\UseRawInputEncoding
\documentclass[twocolumn,superscriptaddress,floatfix, aps, longbibliography, sort]{revtex4}
\usepackage{graphicx,amsfonts,amssymb,amsmath,hyperref,enumerate,footnote}
\usepackage{graphicx}
\usepackage{float}
\usepackage{indentfirst}
\usepackage{verbatim}
\usepackage{xcolor}
\usepackage{tikz}
\usetikzlibrary{quantikz2}
\usetikzlibrary{positioning}
\usepackage{array}
\usepackage[utf8]{inputenc}
\usepackage{mathtools}
\usepackage{blindtext}
\usepackage{pdflscape}
\usepackage{xcolor}
\usepackage{tikz}
\usepackage{algorithm}
\usepackage{algpseudocode}
\usetikzlibrary{arrows.meta,
                chains,
                positioning,
                shapes.geometric}
\usepackage{siunitx}
\tikzstyle{startstop} = [rectangle, rounded corners, minimum width=3cm, minimum height=1cm,text centered, draw=black, fill=red!30]
\tikzstyle{io} = [trapezium, trapezium left angle=70, trapezium right angle=110, minimum width=3cm, minimum height=1cm, text centered, draw=black, fill=blue!30]
\tikzstyle{process} = [rectangle, minimum width=3cm, minimum height=1cm, text centered, draw=black, fill=orange!30]
\tikzstyle{decision} = [diamond, minimum width=3cm, minimum height=1cm, text centered, draw=black, fill=green!30]
\tikzstyle{arrow} = [thick,->,>=stealth]
\newif\ifhyper
\hypertrue
\ifhyper
\hypersetup{
   citecolor = {green},
   colorlinks = {true}, 
   urlcolor = {blue} 
}
\fi

\newcommand{\beq}{\begin{equation}}
\newcommand{\eeq}{\end{equation}}
\newcommand{\beqa}{\begin{eqnarray}}
\newcommand{\eeqa}{\end{eqnarray}}

\def\ket#1{\vert#1\rangle}

\def\Longarrow{\protect\@lra}
\def\@lra{\relbar\joinrel\relbar\joinrel\relbar\joinrel%
          \relbar\joinrel\rightarrow}

\begin{document}

\title{Hacking Cryptographic Protocols with Tensor Network Attacks}

\author{Borja Aizpurua\textsuperscript{*}}
\affiliation{Multiverse Computing, Paseo de Miram\'on 170, E-20014 San Sebasti\'an, Spain}
\affiliation{Department of Basic Sciences, Tecnun - University of Navarra, E-20018 San Sebasti\'an, Spain}

\author{Siddhartha Patra}
\affiliation{Multiverse Computing, Paseo de Miram\'on 170, E-20014 San Sebasti\'an, Spain}
\affiliation{Donostia International Physics Center, Paseo Manuel de Lardizabal 4, E-20018 San Sebasti\'an, Spain}

\author{Josu Etxezarreta Martinez}
\affiliation{Department of Basic Sciences, Tecnun - University of Navarra, E-20018 San Sebasti\'an, Spain}

\author{Rom\'an Or\'us}
\affiliation{Multiverse Computing, Paseo de Miram\'on 170, E-20014 San Sebasti\'an, Spain}
\affiliation{Donostia International Physics Center, Paseo Manuel de Lardizabal 4, E-20018 San Sebasti\'an, Spain}
\affiliation{Ikerbasque Foundation for Science, Maria Diaz de Haro 3, E-48013 Bilbao, Spain}

\thanks{Corresponding author: borja.aizpurua@multiversecomputing.com}

\begin{abstract}
Here we introduce the application of Tensor Networks (TN) to launch attacks on symmetric-key cryptography. Our approaches make use of Matrix Product States (MPS) as well as our recently-introduced Flexible-PEPS Quantum Circuit Simulator (FQCS). We compare these approaches with traditional brute-force attacks and Variational Quantum Attack Algorithm (VQAA) methods also proposed by us. Our benchmarks include the Simplified Data Encryption Standard (S-DES) with 10-bit keys, Simplified Advanced Encryption Standard (S-AES) with 16-bit keys, and Blowfish with 32-bit keys. We find that for small key size, MPS outperforms VQAA and FQCS in both time and average iterations required to recover the key. As key size increases, FQCS becomes more efficient in terms of average iterations compared to VQAA and MPS, while MPS remains the fastest in terms of time. These results highlight the potential of TN methods in advancing quantum cryptanalysis, particularly in optimizing both speed and efficiency. Our results also show that entanglement becomes crucial as key size increases.
\end{abstract}

\maketitle

\section{Introduction}
\label{sec1}

In the digital age, information security has become paramount \cite{nguyen2015survey}, making cryptology an essential discipline in modern society. Cryptology, the science of encryption and decryption, ensures that data remains confidential and intact during transmission. As information has become a critical asset, the ability to protect it against unauthorized access is crucial. This has led to the development of advanced cryptographic protocols and techniques to safeguard sensitive information \cite{AES, RSA,PQC2024}.

With the advent of quantum computing \cite{quantum-computing}, traditional cryptographic methods face unprecedented challenges \cite{scholten2024assessing,PQC2024}. Quantum computers leverage the principles of quantum mechanics to perform computations at speeds unattainable by classical computers \cite{shor, grover}. This capability threatens current cryptographic systems, as quantum algorithms can potentially break widely used encryption protocols much more efficiently than classical methods \cite{20million, grover-attack,PQC2024}. In the Noisy Intermediate-Scale Quantum (NISQ) era, current quantum devices, while powerful, still struggle with errors and qubit coherence limitations \cite{preskill2018quantum,ibmrigetticoherences}. Despite these challenges, significant progress is being made towards realizing the potential of quantum computers. One area of active research is the development of quantum algorithms that can operate effectively within the constraints of NISQ devices. The Variational Quantum Attack Algorithm (VQAA), previously proposed by us, is one such approach that has shown promise in cryptanalysis by optimizing a set of parameters to find encryption keys \cite{EnhancedVQAA, VQAA}.

Moreover, in the NISQ era, simulations can offer reliable benchmarks of noisy quantum computers \cite{patra_ibm_simulation}, which motivates us further to use our quantum-inspired tensor network methods. Such methods have also seen substantial advancements \cite{Or_s_2014}. Tensor networks such as Matrix Product States (MPS) have proven effective in simulating quantum systems \cite{Or_s_2019, Verstraete_2008}. MPS are especially powerful in modeling one-dimensional systems, efficiently capturing their entanglement structure with polynomial resources. This efficiency stands in contrast to the exponential resources required by brute-force methods. The success of MPS in one-dimensional systems has inspired extensions to higher dimensions, such as Projected Entangled Pair States (PEPS), which handle two-dimensional systems and beyond. As an example, we recently introduced the so-called Flexible-PEPS approach \cite{patra2024projectedentangledpairstates}, able to simulate quantum complex systems without a geometry constraint. When simulating quantum computers, the resulting Flexible-PEPS Quantum Circuit Simulator (FQCS) is particularly powerful. 

In this paper we use TNs to launch attacks on cryptographic models, focusing on symmetric-key ciphers. In particular, we make use of variational algorithms for MPS, as well as FQCS simulations of the VQAA approach. Our approaches allow to scale the attacks to large key sizes without the need of quantum hardware \cite{non-orthogonal}, within reasonable memory and time demands.

To be more specific, our approach addresses the known-plaintext problem, where the attacker has access to both the plaintext (also known as a crib) and its encrypted version (ciphertext) \cite{schneier2007applied, book_crypto}, potentially allowing for the revelation of secret keys. The tensor network methods we explore are inspired by the VQAA method \cite{EnhancedVQAA}, which encodes a known ciphertext as the ground state of a classical Hamiltonian. This Hamiltonian represents a cost function, and the method seeks to find this ground state by optimizing a set of variational parameters, ultimately retrieving the secret key. In our current approach, the key is generated either by an MPS (through sampling) or by approximately simulating a Variational Quantum Circuit (VQC) with the tensor network simulator FQCS. Our study focuses on three cryptographic protocols: Simplified Data Encryption Standard (S-DES) with 10-bit keys \cite{sdes, des}, Simplified Advanced Encryption Standard (S-AES) with 16-bit keys \cite{AES}, and Blowfish with 32-bit keys \cite{blowfish}. 

In parallel with quantum and tensor-network-based cryptanalysis, recent cryptographic research has explored biologically inspired and hardware-driven encryption techniques, particularly for multimedia data. For example, chaotic systems and neural models have been employed in image and video encryption using 3D memristive cubic maps with dual discrete memristors \cite{10904857}, 2D Logistic-Rulkov neuron maps \cite{10878977}, and 2D extended Schaffer function maps combined with neural networks \cite{GAO2024520}. While these works target different applications—mainly focusing on data protection in visual and IoT contexts—they underscore the diversity of approaches emerging in modern cryptographic research and the growing importance of alternative computational models.

Recent quantum cryptanalysis efforts have explored both gate-based and annealing-based approaches. For instance, Phab et al. \cite{phab2022first} implemented the Quantum Approximate Optimization Algorithm (QAOA) on the Heys cipher using IBM’s NISQ devices, highlighting the limitations of current quantum circuit depth. Similarly, Pei et al. \cite{10979788} proposed a quantum annealing-based attack using D-Wave hardware on SPN ciphers, formulating the key-recovery problem as a constrained quadratic model. While these works demonstrate early hardware-based feasibility, they target toy ciphers and rely on quantum access. In contrast, our work introduces quantum-inspired tensor network methods—specifically, MPS and FQCS—which enable efficient classical simulation of quantum attacks on realistic symmetric-key protocols (S-DES, S-AES, Blowfish) without requiring access to quantum hardware. Our contributions lie in (i) algorithm design combining tensor networks with variational principles and Hamiltonian simulation, (ii) performance benchmarking across increasing cipher complexity, and (iii) analysis of scalability in entanglement, key size, and time/space cost.

This paper is structured as follows. Section \ref{sec2} provides a detailed overview of the methodologies employed, including the specific tensor network techniques and the setup of our experiments. In Section \ref{sec3}, we present the results of our numerical analysis for S-DES, S-AES, and Blowfish, comparing the performance of MPS, FQCS, VQAA, and brute-force approaches. Finally, Section \ref{sec4} discusses the implications of our findings and potential future improvements together with concluding remarks.

\section{Methods}
\label{sec2}

This section outlines the methodologies used to benchmark Matrix Product States (MPS) and the tensor network-based quantum computer simulator (FQCS) against brute-force attacks and the Variational Quantum Attack Algorithm (VQAA). We detail the implementation, optimization strategies, and experimental setup for each of them in symmetric-key cryptographic protocols.

\subsection{Variational Quantum Attack}
\label{sec2.1}

\begin{figure}
\centering
  \includegraphics[width=0.95\columnwidth]{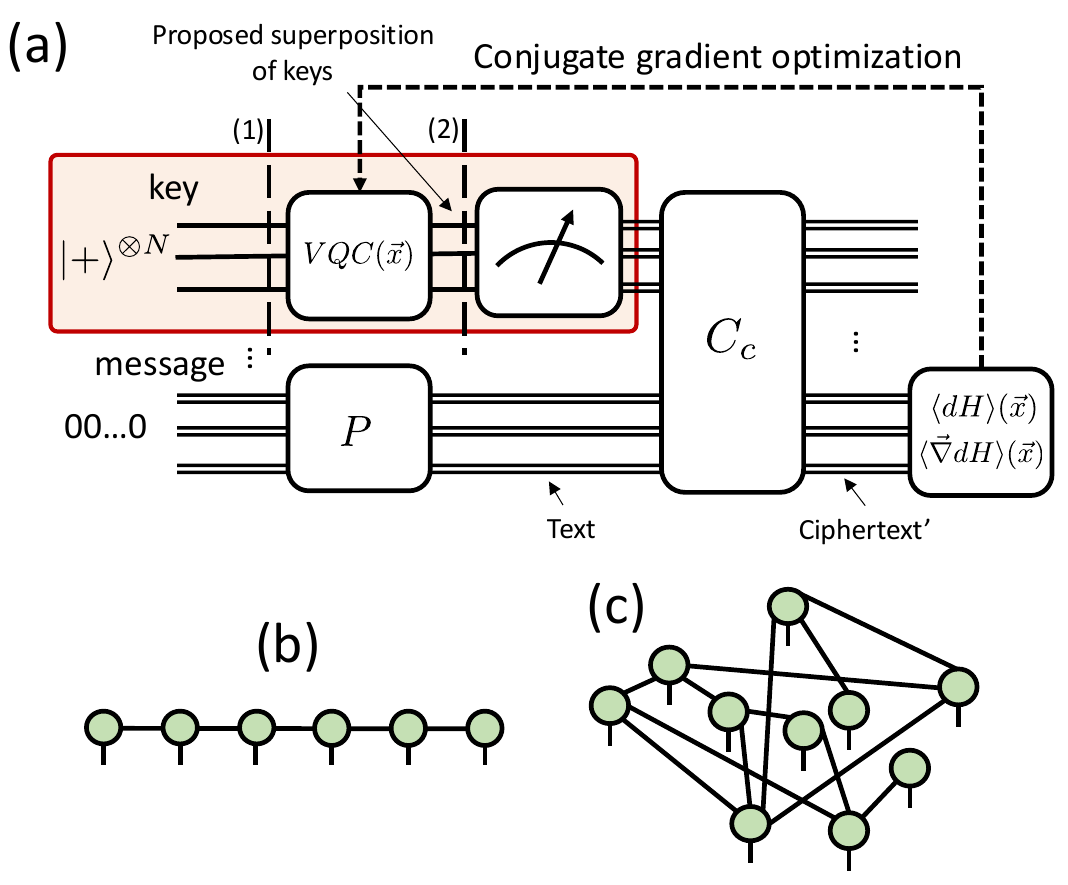}
  \caption{[Color online] (a) Schematic diagram of the improved VQAA. Single lines represent qubits, and double lines represent classical bits. The circuit includes a proposed superposition of keys (left), a parameterized quantum circuit \(VQC(\beta)\), and a conjugate gradient optimization loop (dashed line) that iteratively refines the parameters \(\beta\) to minimize the cost function. The process involves encoding the message, generating a ciphertext, and updating the parameters based on the gradient of the Hamming Distance. The part that runs on a quantum computer is highlighted. (b) MPS of six sites. (c) Arbitrary PEPS without geometry for 10 sites, like the ones that show up in the Flexible-PEPS method.}

  \label{Fig1_VQAA}
\end{figure}

The Variational Quantum Attack Algorithm (VQAA), as developed in Ref. \cite{EnhancedVQAA} and inspired by Ref. \cite{VQAA}, enables the recovery of a cryptographic key given a message (plaintext) and its corresponding encrypted message (ciphertext). As illustrated in Fig.~\ref{Fig1_VQAA}(a), a Variational Quantum Circuit (VQC) is employed to generate the key space and sample potential keys. These keys are subsequently used to encrypt the known plaintext, and the resulting ciphertext is compared to the known ciphertext. Classical optimization techniques are then applied to update the variational parameters of the VQC, and the process is repeated iteratively until the target key is discovered. This method has been demonstrated to be more efficient than traditional brute-force approaches \cite{EnhancedVQAA}.

The VQAA incorporates the use of non-orthogonal states \cite{non-orthogonal}, which involves selecting sets of states that are not perfectly distinct (orthogonal) but are designed to represent different bit configurations with maximal distinguishability. Mathematically, this can be defined as the following minimization problem:
\begin{equation}
\begin{aligned}
 \underset{A\in \mathcal{C}^{2^N\times k}}{\text{argmin}} \quad & \| A^\dagger A - I_{k\times k} \|_F \\
\textrm{s.t.} \quad & a_i^\dagger a_i = 1,\\
\end{aligned}
\end{equation}
where the columns of matrix \( A \), i.e., \( a_i \), represent the \( k \) maximally orthogonal states over \( N \) qubits and \(\|\cdot\|_F\) denotes the Frobenius norm. For simplicity, one of the states is often chosen to be the \(\ket{0}^{\otimes N}\) state. This process requires local (individual) qubit state tomography, which can be performed in parallel for each qubit to obtain the reduced density matrices, from which the most probable state is extracted. Note that in our work, these non-orthogonal states are not physically realized on quantum hardware, but rather classically emulated as part of our simulation framework.

In the context of the VQAA, the \emph{Hamming Distance} has proven to be the most effective cost function among several tested options, including quadratic polynomials, higher-order polynomials, and \( p \)-norms. This metric measures the difference between two binary strings \( a \) and \( b \), specifically the number of bits in which they differ. The Hamming Distance is computed by summing up the bitwise XOR operations \( (a_i \oplus b_i) \):
\begin{equation} \label{eq_dH}
d_H(a,b) = |\{i : a_i \neq b_i\}| = \sum_{i=1}^n (a_i \oplus b_i).
\end{equation}
Moreover, the ansatz for the VQC includes single-qubit unitary rotation gates \( U(\theta, \varphi, \lambda) \) and CNOT gates. 

Concerning the optimization of parameters in the VQC, using the hyperspherical coordinates approach \cite{rotaxis}, the cost function \( f(\vec{x}) \) and original \( n \) Cartesian coordinates define a point \( P \) in a \( (n + 1) \)-dimensional space. The coordinates are then transformed into \( (n+1) \)-dimensional hyperspherical coordinates \( \{\vec{\theta}, r\} \) as follows:
\begin{equation} 
    P = \left[ x_1, x_2, \cdots, x_n, f(\vec{x}) \right] \rightarrow P = \left[ \theta_1, \theta_2, \cdots, \theta_n, r \right].
\end{equation}
Optimization is performed on this new set of coordinates \( \{\vec{\theta}, r\} \) with respect to the reference cost value \( f(\vec{x}) \). Each point is projected from hyperspherical coordinates back into Cartesian coordinates to obtain the new reference value, and then back to hyperspherical coordinates where parameters are updated using the Adam optimizer \cite{Adam}. This transformation strategy helps to mitigate the problem of barren plateaus, which are regions of flat gradients that hinder optimization in high-dimensional spaces. The new point defined by the change \( \{\vec{\theta}, r\} \rightarrow \{\vec{\theta}', r'\} \) after parameter updates is:
\begin{equation}
    P' = \left[ \theta'_1, \theta'_2, \cdots, \theta'_n, r' \right].
\end{equation}
Upon convergence, the final transformation \( \{\vec{\theta}', r'\} \rightarrow \{\vec{x_{\rm c}}, f_c(\vec{x_{\rm c}})\} \) is performed, yielding the point \( P_{\rm c} \):
\begin{equation}
    P_{\rm c} \equiv \left[ x'_1, x'_2, \cdots, x'_n, f'(\vec{x'}) \right].
\end{equation}
This method has shown significant improvements in the number of iterations required for convergence, making it a valuable addition to our variational quantum attack algorithms.

Here we explore also some variations of our original VQAA method. In VQAA we use a variational quantum circuit (\( |\psi(\vec{x})\rangle = U(\vec{x}) |0\rangle^{\otimes N} \)) consisting of single-qubit gates and entangling two-qubit CNOT gates, where \(\vec{x}\) are the parameters. Finally, we sample from this quantum circuit state to recover the key. However, the role of entanglement in key recovery remains unclear. To investigate this, we also adopted a mean-field ansatz, where we use a parameterized quantum circuit but without the use of CNOT gates, i.e. involving only single-qubit gates. This circuit is fully classical, but as we shall see, it may already provide good results for small key sizes. Larger keys, however, benefit more from entangling gates in the circuit. 

In addition, we explored an alternative approach by utilizing a parameterized Hamiltonian:
\begin{eqnarray}
    H(\vec{y}) &=& \sum_i O^{(1)}_i(\vec{y}) +\sum_{ij} O^{(2)}_{ij}(\vec{y}),
\end{eqnarray}
where the one-qubit ($O^{(1)}$) and two-qubit ($O^{(2)}$) terms are exactly those employed in the parametrized quantum circuit as one-qubit and two-qubit gates. The algorithm proceeds through the following steps:
\begin{enumerate}
    \item First, we simulate the ground state of this Hamiltonian by doing imaginary time evolution using our Flexible PEPS tensor network simulator.
    \item We recover a key by sampling from the ground state, which is then used to generate the ciphertext.
    \item Then we optimize the parameters $\vec{y}$ to minimize the distance from the target ciphertext.
\end{enumerate}
This optimization loop is repeated until the target ciphertext is successfully recovered. We call this Hamiltonian based variational quantum algorithm VQAA-h. We find that VQAA-h is comparable when one bit of the key is assigned to one qubit/tensor, i.e., when non-orthogonal states are not used. However, when assigning multiple classical bits to a qubit through non-orthogonal states, the performance of VQAA-h significantly degrades.

As detailed in \cite{EnhancedVQAA} the results for VQAA were obtained using a statevector simulation on a classical computer, specifically utilizing Qiskit's statevector simulator. This simulation computes the exact quantum state vector for the given quantum circuit, which is then used to determine the most probable key. For each qubit, we perform partial trace operations to extract its reduced density matrix, and then calculate fidelities between the quantum state and the possible bit configurations. By analyzing the fidelities for each qubit, the most likely bit values are identified, which are then combined to form a candidate key. This method allows us to recover the key by selecting the configuration with the highest fidelity, ensuring that the most probable key is sampled from the quantum state.

\subsection{MPS Attack}

Our MPS implementation is inspired by Ref.\cite{Han_2018} but is fully reformatted to a new functionality. This process is similar to the one in Subsec. \ref{sec2.1} but with the quantum circuit being replaced by an MPS. Unlike the previous approach, where variational parameters of the VQC are updated to find the key, in the MPS approach, the tensors are updated directly to sample a key closer to the target one.

As usual, we define the MPS with open boundary conditions as:
\begin{equation}
    \left| \psi \right\rangle = \sum_{i_1, i_2, \ldots, i_N} A^{i_1} A^{i_2} \cdots A^{i_N} \left| i_1 i_2 \cdots i_N \right\rangle
\end{equation}
where \( A^{i_k} \) are tensors associated with each site \( k = 1, 2, \cdots, N \), and the indices \( i_k \) represent the physical states at each site, see Fig.~\ref{Fig1_VQAA}(b) for the diagrammatic notation.  Using this, the MPS-based key recovery algorithm is detailed as follows:
\begin{enumerate}    
\item Initialize an MPS with a number of tensors equal to the key size, using a random distribution.
\item Set hyperparameters:
\begin{itemize}
    \item \textbf{BondDim}: Virtual bond dimension of the MPS (e.g., 20).
    \item \textbf{step\_length}: Learning rate for parameter updates (e.g., $10^{-2}$).
    \item \textbf{steps}: Number of updates per tensor (e.g., 1).
    \item \textbf{cutoff\_value}: Threshold for SVD truncation (e.g., $10^{-8}$).
    \item \textbf{reset\_value}: Gradient norm threshold to reset parameters (e.g., 25).
    \item \textbf{temperature}: Controls stochastic acceptance in updates (e.g., 1).
\end{itemize}
\item Left canonicalize the MPS and generate a key sample by sampling the MPS.
\item Calculate the cost function using the Hamming Distance (Eq. \ref{eq_dH}). 
\item Update Sweep: for each tensor from the right-most to the left-most, do
\begin{enumerate}
    \item Merge the current tensor with its neighbor to form a combined matrix, called merged matrix.
    \item Update the merged matrix by applying a small random change, normalize, and then decompose back into the original tensors via SVD.
    \item Left canonicalize the MPS and generate a new key sample.
    \item Calculate the cost function with respect to the target ciphertext.
    \item{If the change is favorable then accept the change. If it is not favorable, accept the change with probability:
        \begin{equation}
            P(\Delta E, T) \propto \exp\left( -\frac{\Delta E}{T} \right), 
        \end{equation}
with $T$ the temperature.} 
    \item Apply the Adam optimizer to update the merged matrix:
    \begin{itemize}
        \item Calculate gradients and update parameters.
        \item Normalize and decompose back into the original tensors via SVD.
    \end{itemize}
    \item Monitor gradient norms to detect local minima; if detected, reset parameters.
\end{enumerate}
\item Right canonicalize the MPS.
\item Repeat the Update Sweep, sweeping from left to right.
\item Continue left-to-right and right-to-left sweeps until the target key is found.
\end{enumerate}

In this approach, the optimization is guided by the cost function, calculated as the Hamming Distance between the generated and target ciphertexts. The optimization process employs a simulated annealing method characterized by state and energy calculations, acceptance criteria based on the Metropolis-Hastings criterion, which allows the algorithm to escape local minima. The Adam optimizer is used to adapt the learning rate for each parameter, based on the first and second moments of the gradients, ensuring efficient convergence to the optimal solution. We observed that moderate choices of bond dimension (1–25) and step sizes in the range $10^{-3}$–$10^{-2}$ generally achieved good performance across all tested ciphers. The optimization was robust to small variations in hyperparameters, and excessive tuning was not required.

\subsection{Flexible-PEPS Attack}
The Flexible-PEPS based Quantum Circuit Simulator (FQCS) is a sophisticated simulation tool \cite{patra2024projectedentangledpairstates}, leveraging TN methods to approximately simulate quantum circuits without an underlying lattice constraint. This section details the design, implementation, and advantages of FQCS in the context of quantum cryptographic analysis.

FQCS is built on the concept of flexible-PEPS, which extends the capabilities of traditional tensor networks. PEPS are a class of quantum many-body states described by tensor networks that generalize Matrix Product States (MPS) from one-dimensional to higher-dimensional systems. Typically, PEPS are defined on regular lattices, such as square or cubic grids. However, FQCS employs a flexible geometry approach, allowing the tensor network geometry to adapt dynamically to the system's correlation structure by deleting less correlated connections, see Fig.~\ref{Fig1_VQAA}(c). This adaptability is particularly advantageous for simulating quantum circuits with long-range random interactions and dense connections. FQCS uses a cut-off parameter, $\kappa$, which represents the maximum vertex degree in the tensor network and controls the computational complexity. By enforcing a tunable vertex degree limit, FQCS can manage the exponential growth in computational resources typically associated with densely connected graphs. This is achieved through an edge-deletion rule based on bond entanglement entropy (BEE), which ensures that the network retains the most significant correlations while discarding less relevant ones.

A detailed step-by-step flowchart of the FQCS simulation algorithm is provided in Appendix~\ref{app:flowchart}, outlining the sequence of gate application, entanglement update, and truncation procedures.

The flexible-PEPS approach in FQCS offers several significant advantages. First,  FQCS can efficiently scale to simulate large quantum systems by dynamically adjusting the network's geometry keeping the computational complexity in check. This scalability is crucial for cryptographic applications that require the simulation of circuits with large key sizes. Second, by using a flexible geometry, FQCS reduces the memory and time requirements compared to fixed-geometry PEPS. This efficiency enables the simulation of complex quantum algorithms that would otherwise be infeasible. And third, the edge-deletion strategy based on BEE ensures that the essential quantum correlations are preserved, maintaining the accuracy of the simulation without using any SWAP gates. This is particularly important for cryptographic analyses, where precise modeling of quantum systems is necessary.

The main tunable parameters for FQCS include the same as in the MPS attack, adding \texttt{number\_layers} (typically 2) and \texttt{number\_qubits} (e.g., 16) for the circuit simulation. As with the MPS attack, performance was robust across reasonable values, with convergence most sensitive to the number of layers and the reset threshold. Unlike in MPS, we did not vary the physical bond dimension extensively in FQCS, as it mostly affects the precision of tensor contractions. With relatively shallow circuits and few qubits, we observed minimal performance differences across bond dimension settings.

Our approach to use FQCS to device a cyberattack is simple: we'll use it to simulate classically the VQAA attack described in Subsec. \ref{sec2.1} and shown in Fig.~\ref{Fig1_VQAA}. As such, we know that the simulation will break down for quantum circuits involving a very large amount of entanglement. But we also know that the classical simulation of the quantum circuit can be a remarkably good, if not the best, classical algorithm for such attacks. We will be benchmarking against traditional brute-force attacks and other quantum-inspired techniques. 

Although FQCS is designed as a classical simulation framework, the quantum circuits it simulates consist entirely of standard one- and two-qubit gates, which are natively supported by current quantum hardware. This makes FQCS-generated circuits directly implementable on gate-based quantum platforms. However, the depth and entanglement of the circuits must be carefully managed to remain within the limits imposed by decoherence and gate noise. In this context, FQCS serves as a valuable tool for prototyping and benchmarking variational quantum attacks prior to deployment on real devices.

\section{Results}
\label{sec3}

In this section, we present the results of our cryptanalysis using tensor network methods on three symmetric-key ciphers. We benchmark these methods against traditional brute-force attacks and the VQAA to evaluate their efficiency and effectiveness in key recovery. We begin with S-DES, the most extensively studied cipher in our experiments, followed by S-AES and Blowfish. To ensure statistical significance, all reported averages are computed over a large number of independent attack attempts (e.g., 200 for S-DES and 100 for S-AES and Blowfish), each with randomized key and plaintext-ciphertext pairs.

\subsection{Simplified-Data Encryption Standard (S-DES)}

S-DES, or Simplified Data Encryption Standard is a cryptographic protocol designed to provide a basic understanding of symmetric-key cryptography. It operates on 8-bit blocks of data using a 10-bit key, employing a series of permutation and substitution techniques to transform plaintext into ciphertext and vice versa. This makes S-DES suitable for educational purposes and basic encryption tasks, providing a foundational understanding of more complex cryptographic protocols like the Data Encryption Standard (DES).

\begin{figure*}[htbp]
    \centering
    \begin{minipage}[!htbp]{0.45\textwidth}
        \centering
        \includegraphics[width=\textwidth]{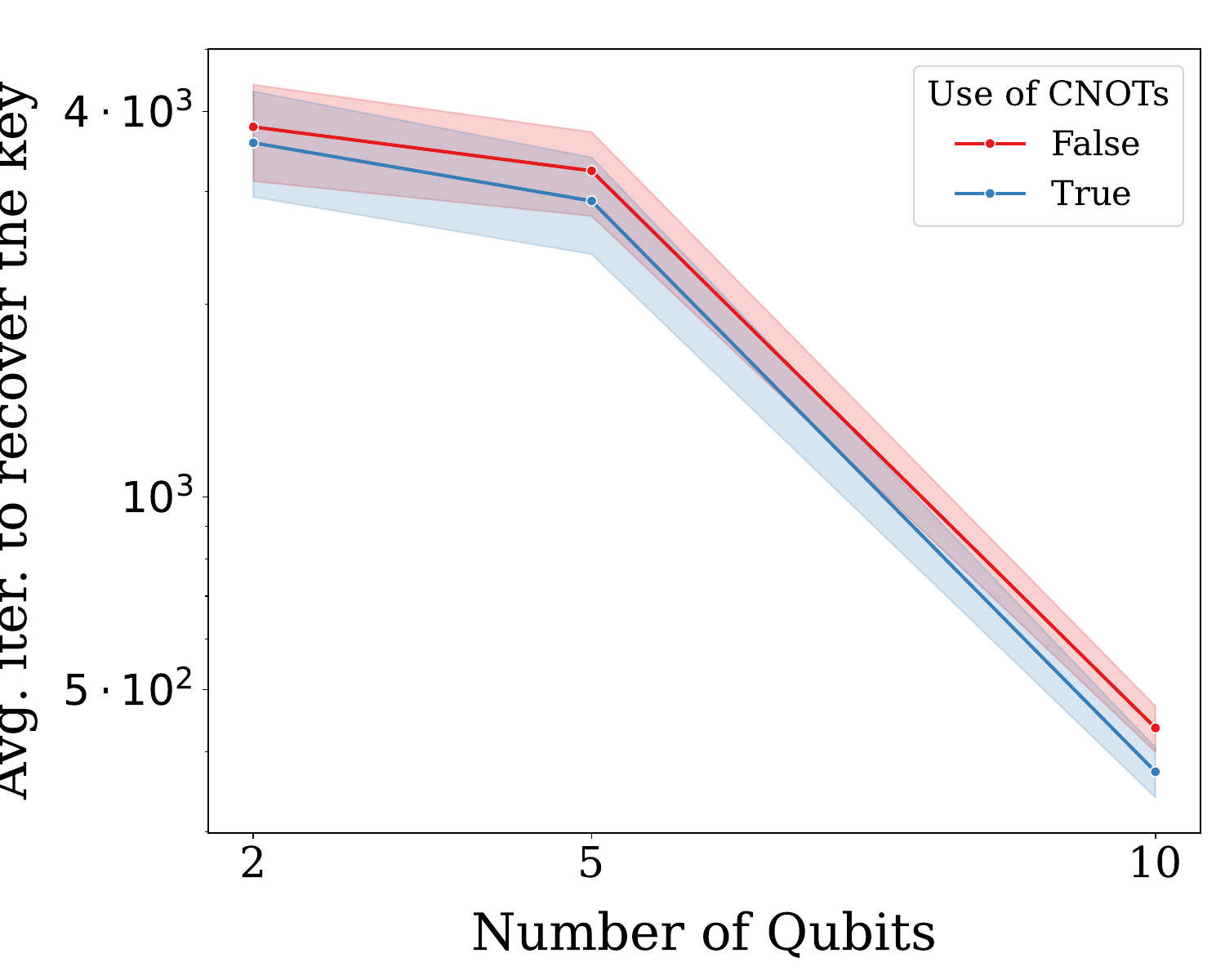}
        \caption{[Color online] Impact of CNOT gates on the average number of iterations to recover the key in S-DES as a function of the number of qubits.}
        \label{Fig2_Meas_cnot}
    \end{minipage}%
    \hfill
    \begin{minipage}[!htbp]{0.45\textwidth}
        \centering
        \includegraphics[width=\textwidth]{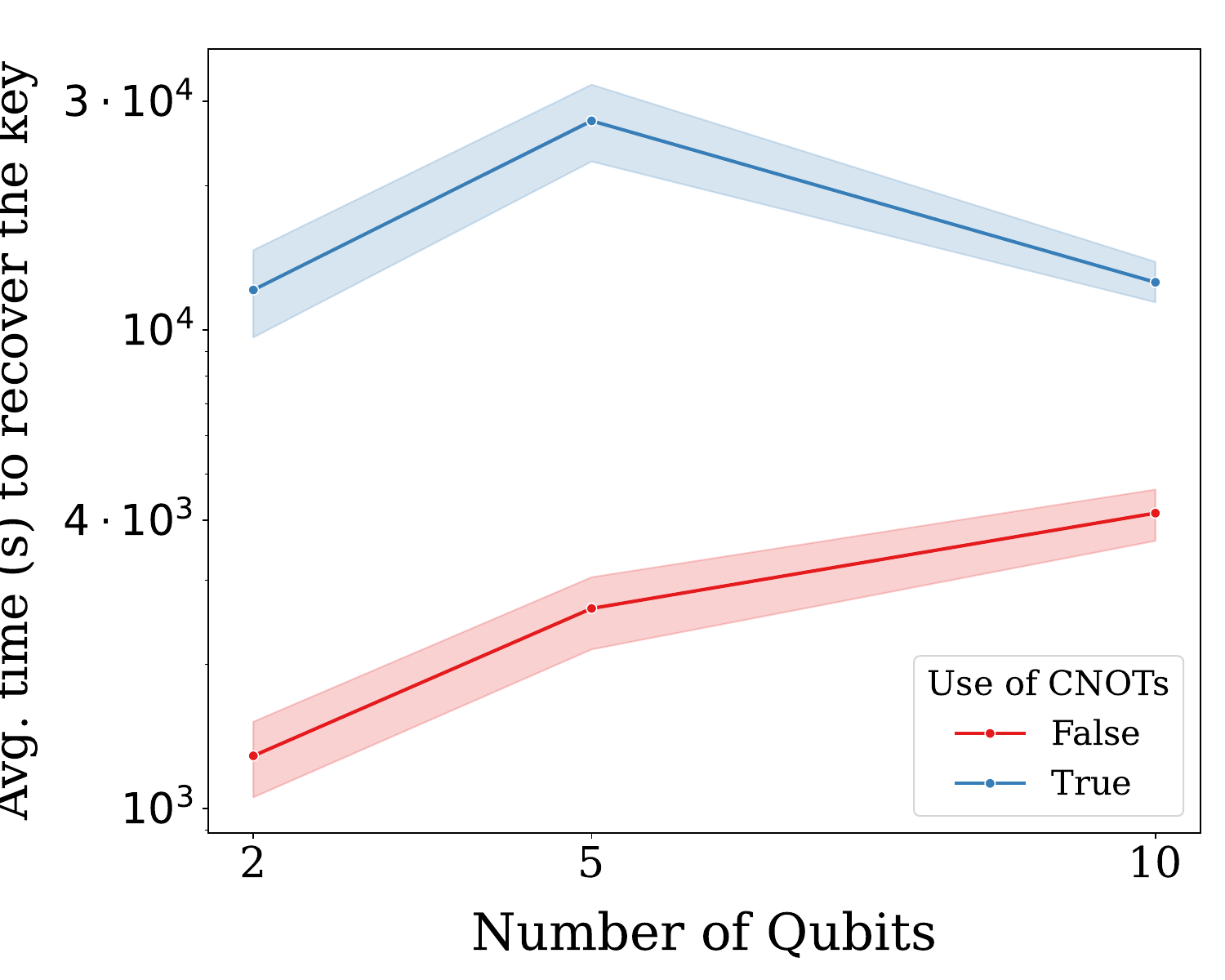}
        \caption{[Color online] Impact of CNOT gates on the elapsed time (seconds) to recover the key in S-DES as a function of the number of qubits.}
        \label{Fig3_Time_cnot}
    \end{minipage}
    
    \vspace{0.5cm} 

    \begin{minipage}[!htbp]{0.45\textwidth}
        \centering
        \includegraphics[width=\textwidth]{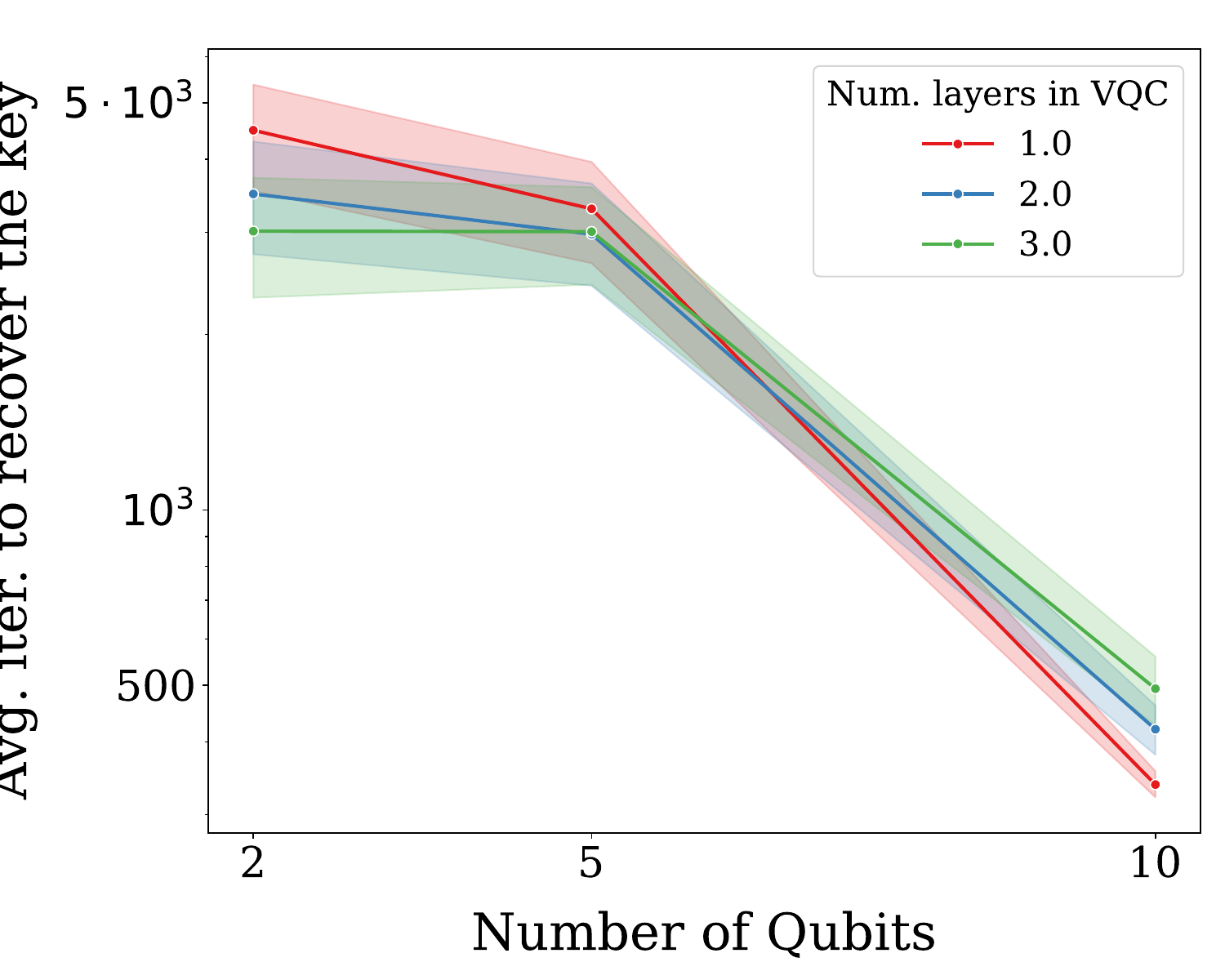}
        \caption{[Color online] Impact of the number of layers in the VQC on the average number of iterations to recover the key in S-DES as a function of the number of qubits.}
        \label{Fig4_Meas_layers}
    \end{minipage}%
    \hfill
    \begin{minipage}[!htbp]{0.45\textwidth}
        \centering
        \includegraphics[width=\textwidth]{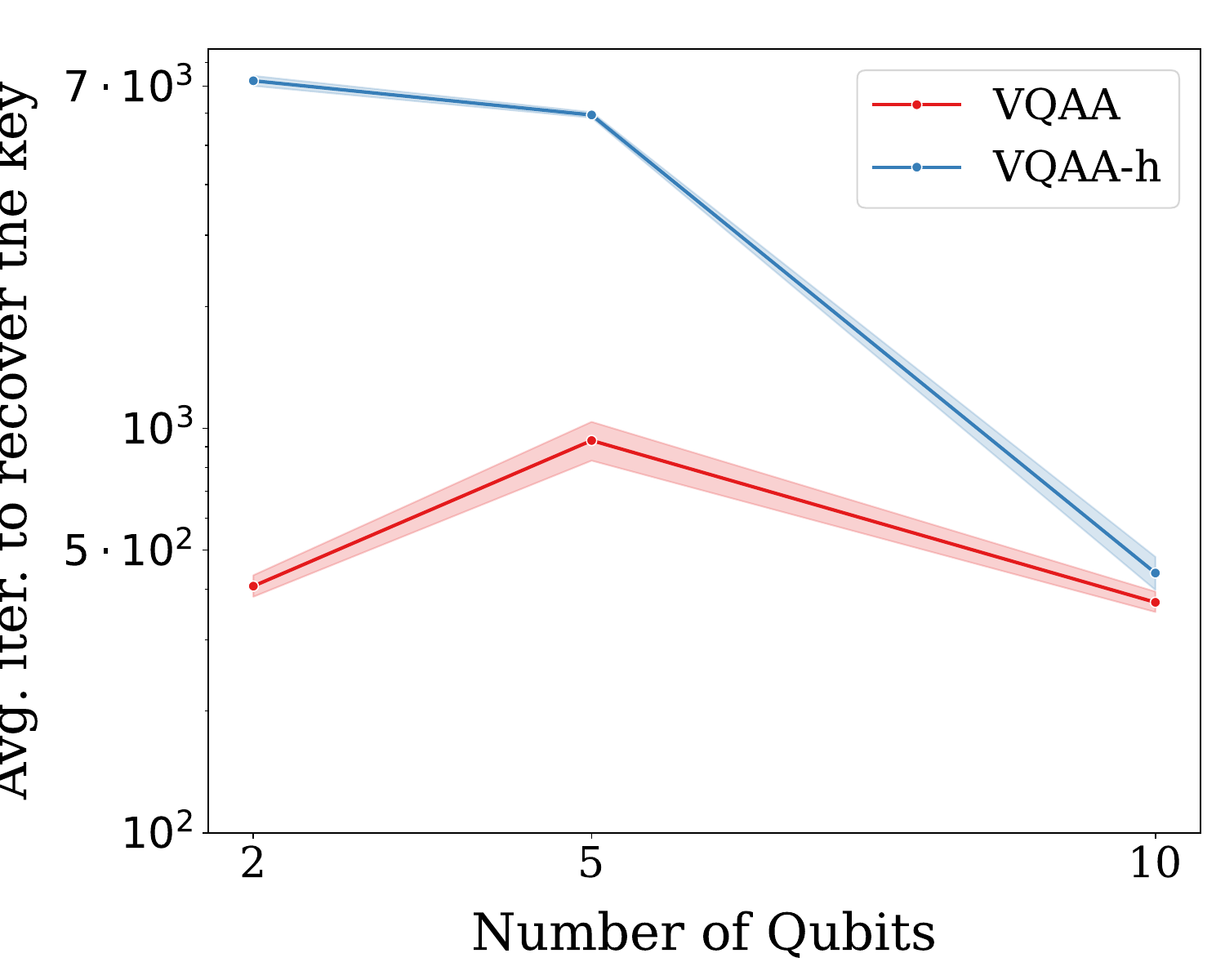}
        \caption{[Color online] Impact of VQAA and VQAA-h (Hamiltonian) on the average number of iterations to recover the key in S-DES as a function of the number of qubits.}
        \label{Fig5_Meas_uhu}
    \end{minipage}
\end{figure*}

A brute-force attack, requiring an average of 512 iterations, served as a baseline for comparison. In a brute-force attack, every possible key is tried until the correct one is found. Given that S-DES uses a 10-bit key, there are $2^{10} = 1024$ possible keys, and on average, half of these (512) would need to be tested to find the correct key. The brute-force approach took 2.51 seconds in our tests, which involved generating all possible keys using a Python script and encrypting the known plaintext with them. This time could be further reduced with dedicated hardware designed for such tasks.

For the FQCS, we conducted an extensive analysis considering various factors such as the number of samples to generate when deciding a key, use of non-orthogonal states or not, number of tensors (qubits), use of CNOT gates or not, and the structure of the quantum circuit (using VQAA or VQAA-h). We also tuned algorithm hyperparameters to optimize performance.

The impact of CNOT gates on the average number of iterations to recover the key is shown in Fig.~\ref{Fig2_Meas_cnot}. the figure illustrates the effect of using CNOT gates on the average number of iterations required to recover the key as a function of the number of qubits. The results show that incorporating CNOT gates slightly enhances performance, reducing the number of iterations needed. The impact of CNOT gates on elapsed time is shown in Fig.~\ref{Fig3_Time_cnot}. The results show that implementations using CNOT gates are significantly slower. There is an increase in time from 2 to 5 qubits, followed by a decrease from 5 to 10 qubits, indicating greater efficiency of the algorithm with 10 qubits. The impact of the number of layers in the VQC on the average number of iterations to recover the key is discussed in  Fig.~\ref{Fig4_Meas_layers}. The results show that having more layers improves performance with fewer qubits but has a diminishing effect as the number of qubits increases. This suggests that with fewer qubits, additional layers provide more variational parameters to explore the key space efficiently, while with more qubits, the number of layers becomes less important. Last but not least, the impact of VQAA and VQAA-h (Hamiltonian) on the average number of iterations to recover the key is shown in Fig.~\ref{Fig5_Meas_uhu}. The results show that while the performances of VQAA and VQAA-h are similar for systems with 10 qubits, VQAA is significantly more efficient for smaller systems with fewer qubits. This difference in performance might be due to the non-orthogonal states not favoring the Hamiltonian-based approach in VQAA-h for smaller qubit numbers.

Concerning the MPS approach, we found that using a bond dimension of 1 (product state) yielded very efficient results, even though this implies a product state with zero entanglement. As shown in Table~\ref{table_sdes}, this approach is more effective than other approaches in terms of the number of iterations it takes to recover the key, and is also much faster (except brute-force), as the algorithm updates the tensors directly.

The most effective configuration of the FQCS method used 10 qubits, without CNOTs (no entanglement), 1 layer, and sampled only once (the generated sample was considered as the key directly). This configuration involved the VQAA-h (Hamiltonian) approach where we sample from the simulated ground state \( |\psi(\vec{y})\rangle \) of a parameterized Hamiltonian \( H(\vec{y}) \), and did not involve non-orthogonal states. While this was the best configuration of FQCS, it was not the most efficient approach overall.

The VQAA approach was utilized as described in \cite{EnhancedVQAA}. The VQAA results, with an average of 238 iterations and 232.2 seconds, were obtained by simulating 5 qubits and encoding 2 bits in each qubit using non-orthogonal states. This method leverages the variational quantum circuit's ability to explore the key space efficiently by optimizing the circuit parameters to recover the key.

\begin{table}[!htbp]
\centering
\begin{tabular}{|c|c|c|}
\hline
\textbf{Method} & \textbf{Average Iterations} & \textbf{Average Time (s)} \\ \hline
Brute-force & 512 & 2.51 \\ \hline
VQAA & 238 & 232.2 \\ \hline
FQCS & 221.9 & 145.4 \\ \hline
MPS & 203.2 & 11.83 \\ \hline
\end{tabular}
\caption{Comparison of average number of iterations and time required to recover 200 keys by various cryptanalysis methods on the S-DES algorithm.}
\label{table_sdes}
\end{table}

\begin{figure}[!htbp]
    \centering
    \includegraphics[width=0.95\columnwidth]{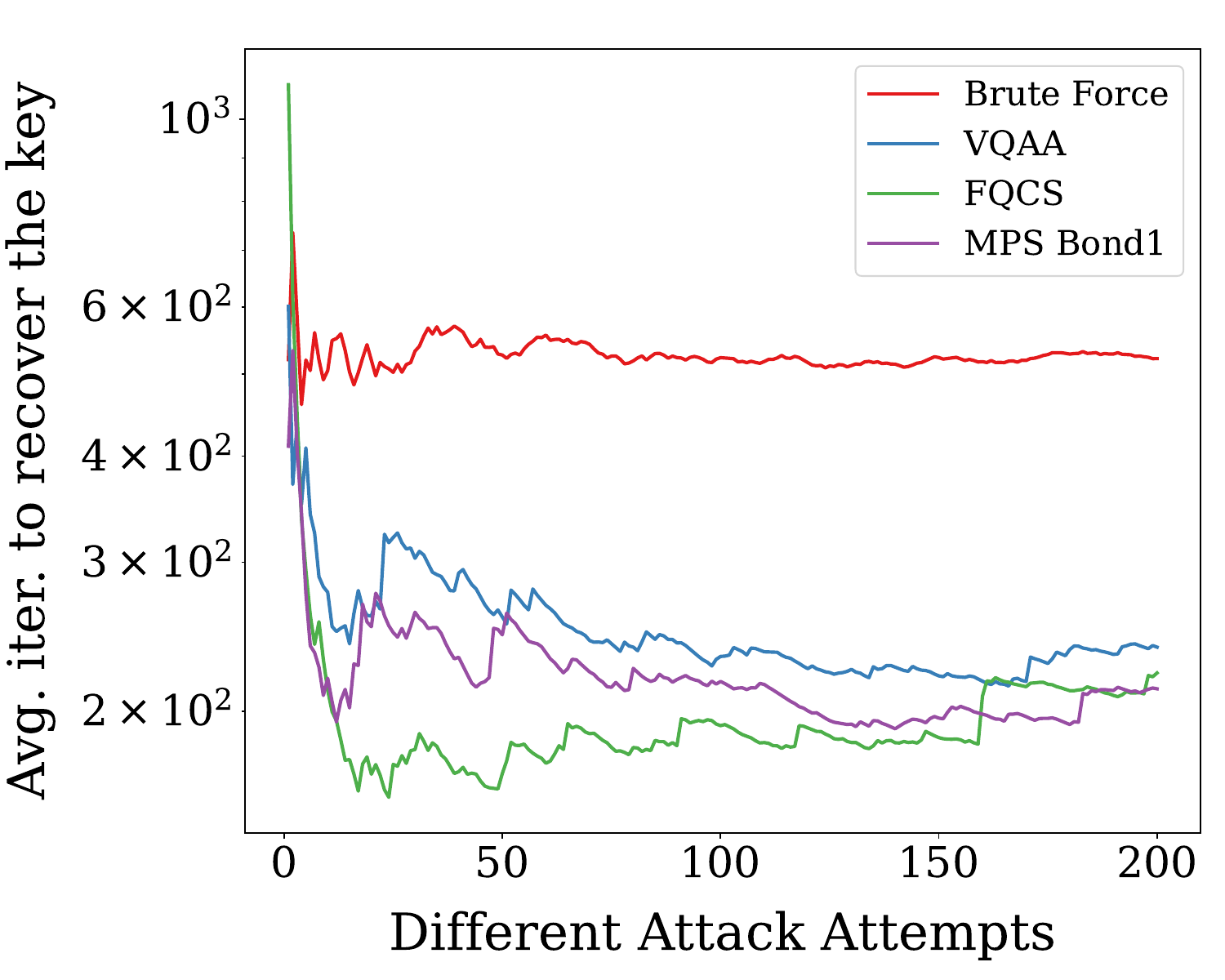}
    \caption{[Color online] Performance comparison of different cryptanalysis methods on S-DES. The plot illustrates the average number of iterations required to recover the key across 200 different attack attempts (single, independent execution of the cryptanalysis algorithm using a unique key and a corresponding plaintext-ciphertext pair) using Brute Force, VQAA, FQCS, and MPS with Bond Dimension 1.}
    \label{Fig6_results_sdes}
\end{figure}

Our extensive analysis of S-DES cryptanalysis is summarized in Fig.~\ref{Fig6_results_sdes} and reveals that MPS with a bond dimension of 1 (product state) outperforms TN and Quantum based methods in terms of both iterations and computational time, effectively serving as a mean-field approach, although brute-force has the lowest runtime overall. This result indicates that entanglement is not necessary to recover the key efficiently in the case of S-DES, likely due to the simplicity and small key size of the cipher. However, this may not hold true for more complex ciphers with larger key sizes, which will be examined in the following sections. The FQCS approach, while efficient, shows significant variability based on the configuration of qubits, layers, and gate structures. 

\subsection{Simplified Advanced Encryption Standard (S-AES)}

The Simplified Advanced Encryption Standard (S-AES) \cite{S-AES} is an educational adaptation of the widely-used Advanced Encryption Standard (AES) algorithm, designed to aid in understanding the complexities of AES. S-AES simplifies key components of the encryption process, making it accessible for learning and illustrative purposes. It operates on smaller data blocks, typically 16 bits, and uses shorter key lengths compared to AES, typically 16 bits. S-AES employs substitution-permutation networks, a core concept in modern symmetric-key cryptography, to transform plaintext into ciphertext and vice versa. By reducing the number of rounds and employing straightforward operations, S-AES offers a practical and comprehensible entry point to study encryption algorithms.

We benchmarked the same four cryptanalysis methods as with S-DES: brute-force, VQAA, MPS, and FQCS. For S-AES, we conducted 100 runs for each method, using different plaintext-ciphertext pairs in each run to evaluate the average number of iterations needed to recover the correct key and the computational time for each iteration.

A brute-force attack, requiring an average of 32,768 iterations, served as a baseline for comparison. In a brute-force attack, every possible key is tried until the correct one is found. Given that S-AES uses a 16-bit key, there are $2^{16} = 65,536$ possible keys, and on average, half of these (32,768) would need to be tested to find the correct key. The brute-force approach took less than 14 seconds in our tests, which involved generating all possible keys using a Python script and encrypting the known plaintext with them. This time could be further reduced with dedicated hardware designed for such tasks.

For the FQCS, we conducted a similar analysis to the one in S-DES, considering factors such as the number of samples to generate when deciding a key, use of non-orthogonal states, different numbers of tensors (qubits), use of CNOT gates, and the structure of the quantum circuit (VQAA or VQAA-h).

The results for S-AES are summarized in Table~\ref{table_saes}. Unlike S-DES, the FQCS was more effective in terms of the average number of iterations it takes to recover the key. The best FQCS configuration used 16 qubits, 1 sample, no CNOTs (mean-field ansatz), 1 layer, and the VQAA-h approach. Again, including CNOTs improves average number of iterations a little bit but in terms of time (due to simulation runtime) it's too slow.

\begin{table}
\centering
\begin{tabular}{|c|c|c|}
\hline
\textbf{Method} & \textbf{Average Iterations} & \textbf{Average Time (s)} \\ \hline
Brute-force & 32,768 & 13.48 \\ \hline
VQAA & 23,584 & 8,241 \\ \hline
FQCS & 13,067.1 & 7,231 \\ \hline
MPS & 31,303 & 1,162 \\ \hline
\end{tabular}
\caption{Comparison of average number of iterations and time required to recover 100 keys by various cryptanalysis methods on the S-AES algorithm.}
\label{table_saes}
\end{table}

The VQAA approach also followed the methodology outlined in \cite{EnhancedVQAA}. The results, with an average of 23,584 iterations and 8,241 seconds, were achieved by simulating 4 qubits and encoding 4 bits in each qubit using non-orthogonal states. This configuration allowed the VQAA to handle the increased complexity and key size of S-AES, demonstrating its capability to adapt to more challenging cryptographic problems.

Again we found that using a bond dimension of 1 yielded faster results due to the algorithm's simplicity in the MPS approach, but it was no longer more effective than FQCS in terms of the average number of iterations needed to recover the key. We believe the downgrade in performance could be due to over-parametrization (too many weights to optimize) if bond dimension is greater than 1, or lack of parameters in the case of bond dimension 1. We also tested higher bond dimensions and observed that while they generally reduced the number of iterations required, they increased runtime due to the added computational overhead. As such, a bond dimension of 1 provided the best trade-off between accuracy and speed for small key sizes.

\begin{figure}
    \centering
    \includegraphics[width=0.95\columnwidth]{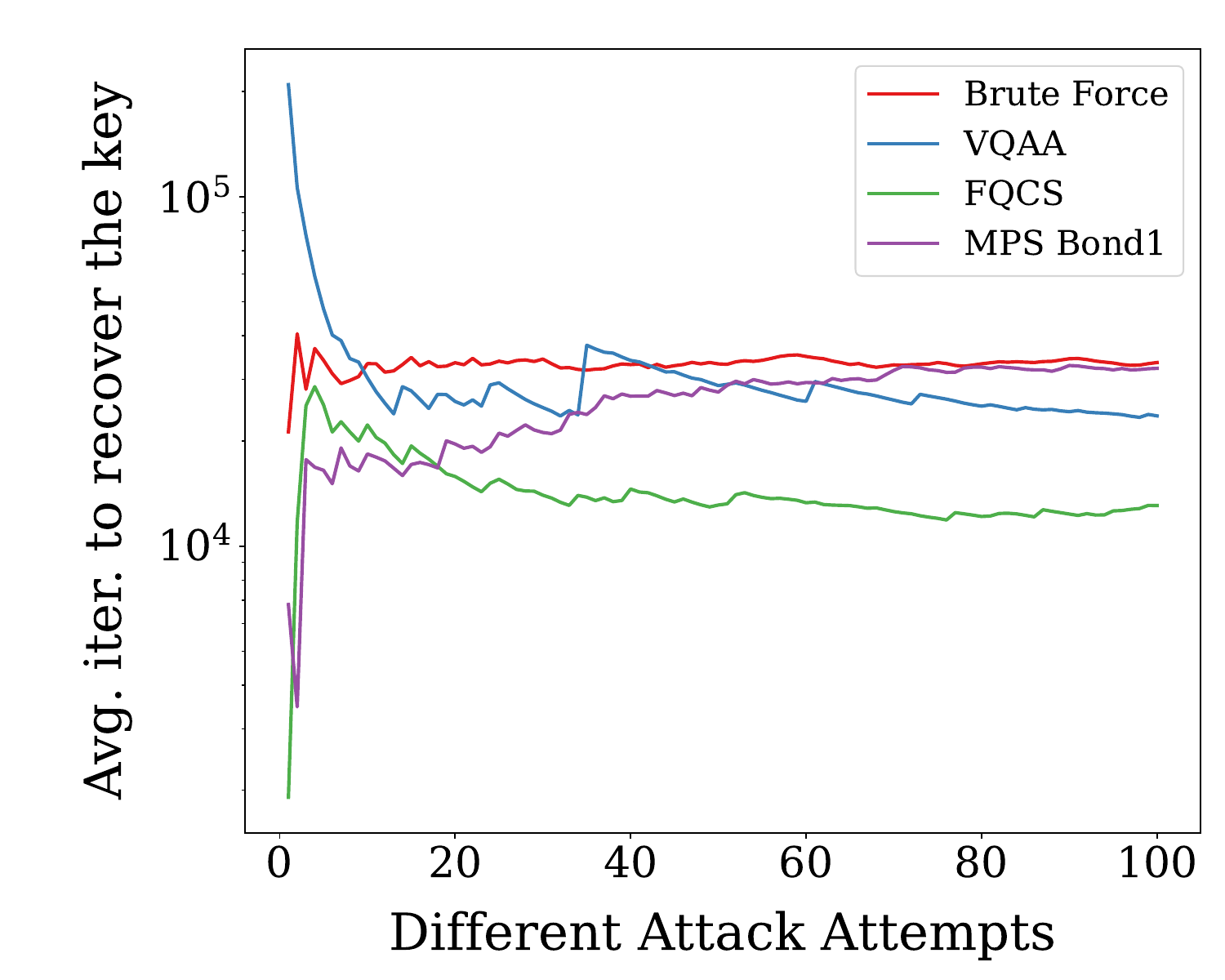}
    \caption{[Color online] Performance comparison of different cryptanalysis methods on S-AES. The plot illustrates the average number of iterations required to recover the key across 100 different attack attempts using Brute Force, VQAA, FQCS, and MPS with Bond Dimension 1.}
    \label{Fig7_results_saes}
\end{figure}

Our analysis of S-AES cryptanalysis, summarized in Fig.~\ref{Fig7_results_saes}, reveals that while MPS offers faster computation times compared to FQCS and VQAA due to its algorithmic simplicity, it is not as efficient as brute-force methods in terms of raw speed. Among quantum-inspired approaches, FQCS proves more effective in terms of the average number of iterations needed to recover the key, particularly when configured without CNOTs—effectively serving as a mean-field ansatz. This highlights the importance of optimizing circuit configurations for practical quantum cryptanalysis.

A direct comparison between MPS and VQAA-h further clarifies this trade-off. As shown in Fig.~\ref{Fig7_results_saes}, VQAA-h outperforms MPS in terms of average iterations required to recover the key. However, MPS remains significantly faster overall due to its lower per-iteration cost and simpler tensor updates. This reflects a core distinction: VQAA-h, simulated via FQCS, accesses a richer variational space incorporating entanglement, but at higher computational expense. In contrast, MPS—especially at low bond dimension—offers rapid convergence with reduced expressivity. As cipher complexity and key size increase, we anticipate that the advantages of VQAA-h may become more pronounced, provided sufficient simulation resources are available.

\subsection{Blowfish}

Blowfish is a symmetric-key block cipher encryption algorithm \cite{blowfish}, known for its simplicity and efficiency. It operates on fixed-size blocks of data and supports key lengths ranging from 32 bits to 448 bits, making it adaptable to various security requirements. Its key setup phase is notably fast, enabling rapid encryption and decryption processes. Despite its age, Blowfish remains widely used and respected due to its robust security features and speed. Its open design and absence of any licensing restrictions have contributed to its popularity in both commercial and open-source applications. The algorithm's resilience against various cryptanalytic attacks has solidified its reputation as a reliable choice for secure data encryption.

Remarkably, no effective cryptanalysis has been found to date for Blowfish, with brute-force attacks being the standard method. Although the cipher is believed to be weak against birthday attacks, these are also brute-force collision attacks based on the birthday paradox \cite{birthdayattack}.

Given our previous experience with Blowfish in Ref. \cite{EnhancedVQAA}, instead of attempting a full attack on the 32-bit key, we employ a hybrid approach. We fix the first 8 bits of the key to the correct values, reducing the search space to the remaining 24 bits. This approach simulates running $2^8 = 256$ configurations in parallel in few-qubit quantum processors. Our analysis used 6 qubits, implying the need for 256 independent 6-qubit quantum processors, which is realistic with current NISQ technology that handles thousands of qubits, such as those using neutral atoms \cite{atom}. In this approach, a brute-force attack searching $2^{24} = 16,777,216$ possible keys would require on average half of these (8,388,608) to be tested to find the correct key.

Similarly, we benchmarked the performance of the same four different cryptanalysis methods
over 100 runs. The results for Blowfish are summarized in Table~\ref{table_blowfish}.

\begin{table}
\centering
\begin{tabular}{|c|c|c|}
\hline
\textbf{Method} & \textbf{Average Iterations} & \textbf{Average Time (s)} \\ \hline
Brute-force & 8,388,608 & 15,558 \\ \hline
VQAA & 4,390,952 & 86,051 \\ \hline
FQCS & 10,647,241 & 351,177 \\ \hline
MPS & 14,844,610 & 89,914 \\ \hline
\end{tabular}
\caption{Comparison of average number of iterations required to recover 100 keys and the time required for each of them by various cryptanalysis methods on the Blowfish algorithm.}
\label{table_blowfish}
\end{table}

Unlike previous algorithms such as S-DES and S-AES, the VQAA approach provided the most effective results in terms of the number of iterations required to recover the key, though brute-force was still faster in terms of computation time. Specifically, the VQAA method, as detailed in \cite{EnhancedVQAA}, utilized CNOT gates and a 6-qubit simulation where 4 bits were encoded into each qubit using non-orthogonal states. This configuration resulted in an average of 4.3 million iterations and 86,000 seconds, underscoring the increasing importance of CNOTs and entanglement as the complexity of the cryptographic algorithm scales.

For the FQCS method, we simulated 24 tensors/qubits, following a configuration with 1 layer, the VQAA approach, 1 sample, and no CNOT gates. This setup, however, proved to be quite slow, with an average of 10.6 million iterations and a significantly longer runtime (351,177 seconds). The lack of CNOTs and the use of a low bond dimension in this configuration limited its effectiveness. It is likely that a more expensive FQCS setup, incorporating CNOTs and higher bond dimensions, could improve the results, but such configurations would require significantly more runtime. The MPS approach, while faster in some scenarios, still failed to be efficient enough to pose a significant threat to the cryptographic security of Blowfish. The results for the MPS method showed an average of 14.8 million iterations and 89,914 seconds, indicating that despite its speed, it is not as effective as VQAA in terms of iteration count.

\begin{figure}
    \centering
    \includegraphics[width=0.95\columnwidth]{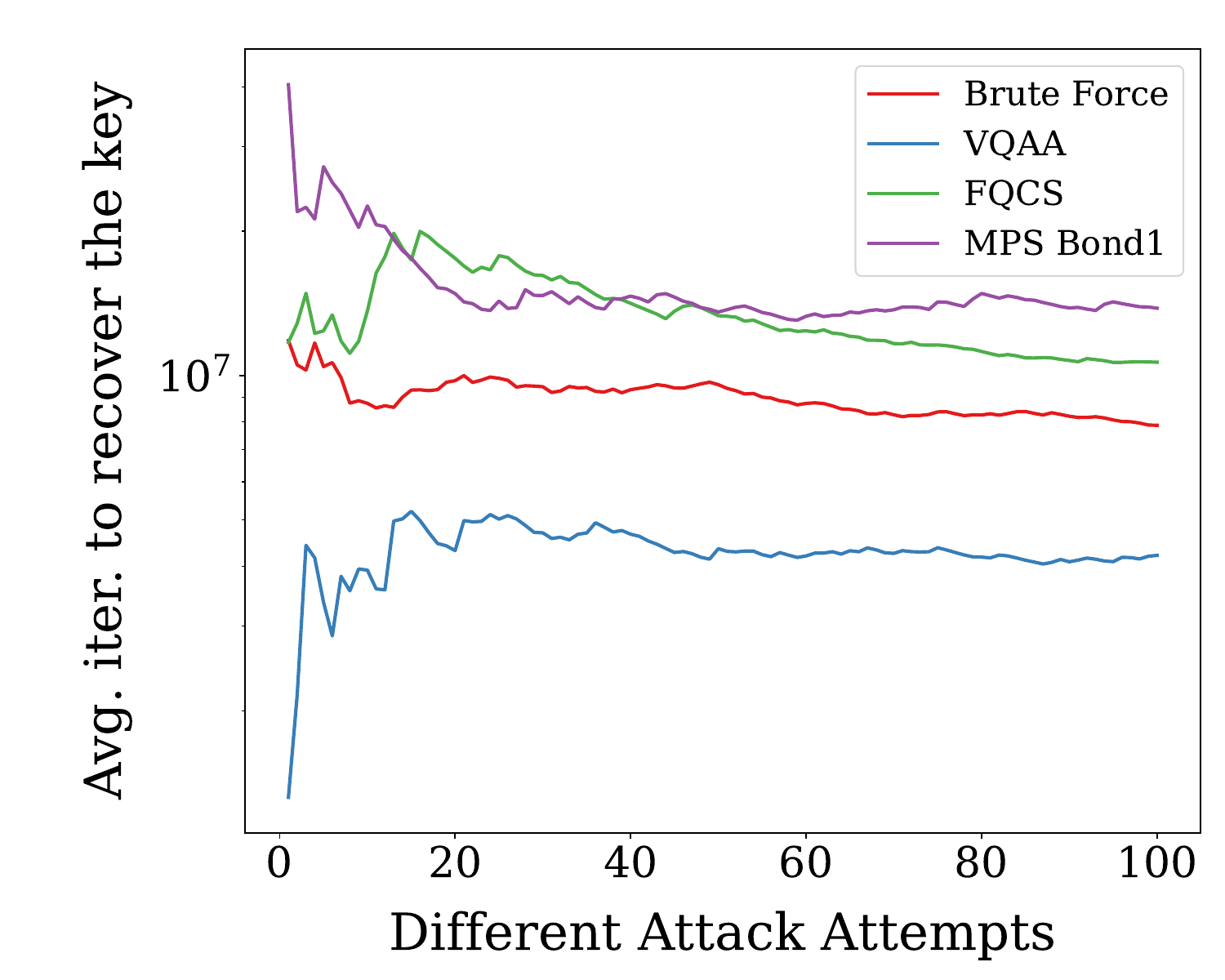}
    \caption{[Color online] Performance comparison of different cryptanalysis methods on Blowfish. The plot illustrates the average number of iterations required to recover the key across 100 different attack attempts using Brute Force, VQAA, FQCS, and MPS with Bond Dimension 1.}
    \label{Fig8_results_blowfish}
\end{figure}

Overall, as it can be seen in Fig.~\ref{Fig8_results_blowfish}, these findings suggest that while VQAA remains a potent method for cryptanalysis, especially as algorithm complexity increases, the FQCS approach may require further optimization, particularly when applied to more complex algorithms like Blowfish. The MPS method, although quick, might lack the depth required to tackle such sophisticated cryptographic challenges.

For the MPS approach, the per-sweep time complexity is $\mathcal{O}(D^3)$ and the space complexity is $\mathcal{O}(D^2)$, where $D$ is the bond dimension. While small bond dimensions (e.g., $D=1$) are sufficient for simple ciphers like S-DES, larger values of $D$ are needed to capture more entanglement in complex ciphers, which significantly increases runtime and memory requirements.

The time complexity of the FQCS algorithm is $\mathcal{O}(D^{2\kappa-1})$, and space complexity $\mathcal{O}(D^{\kappa})$, where $\kappa$ is the max vertex degree. Although FQCS benefits from entanglement pruning, high $\kappa$ values and deep circuits for 64- or 128-bit keys demand large memory and compute resources. In practice, VQAA-h offers better key recovery with fewer iterations for intermediate-size ciphers (e.g., S-AES), but the higher computational overhead makes MPS more practical for quick attacks on simpler protocols (e.g., S-DES).

VQAA simulations, based on full statevector backends, have exponential time and space complexity $\mathcal{O}(2^N)$, limiting practical runs to fewer than 25 qubits. In contrast, brute-force attacks scale as $\mathcal{O}(2^k)$ in time and remain infeasible beyond 64-bit keys.

Overall, while all methods face scaling challenges, tensor network approaches (MPS, FQCS) offer more tractable classical alternatives for intermediate key lengths. Further optimization or parallelization would be required for full 64/128-bit cryptanalysis.

\section{Conclusions}
\label{sec4}

In this study, we analyzed the performance of four cryptanalysis methods—brute-force, Variational Quantum Attack Algorithm (VQAA), Matrix Product States (MPS), and the Flexible-PEPS based Quantum Circuit Simulator (FQCS)—on three symmetric-key ciphers: Simplified Data Encryption Standard (S-DES), Simplified Advanced Encryption Standard (S-AES), and Blowfish. Each method was evaluated based on the average number of iterations needed to recover the correct key and the computational time required for each iteration.

For S-DES, the results indicated that MPS with a bond dimension of 1 (product state) performed remarkably well, benefiting from the simplicity of the cipher and the limited entanglement required. This low-complexity setup allowed MPS to achieve high efficiency in both convergence and runtime. In contrast, for the more intricate S-AES cipher, FQCS showed stronger performance in terms of average iterations, despite longer runtimes. The ability of FQCS to approximate more expressive circuit structures—while remaining classically executable—proved beneficial in this intermediate regime. Finally, for Blowfish, the VQAA method, incorporating entangling CNOT gates and simulated over six qubits, achieved the best key recovery rates. This reflects the increasing relevance of quantum resources as cipher complexity scales.

Overall, these findings suggest that the balance between algorithmic expressivity and computational cost plays a critical role in cryptanalysis. While brute-force attacks remain effective for small key spaces, more sophisticated methods such as MPS, FQCS, and VQAA offer advantages in scalability and search space exploration. The progression of results across the three ciphers hints at a gradual transition: as key size and internal cipher structure become more complex, the effectiveness of quantum-inspired and tensor-based methods becomes increasingly evident.

This evolution is summarized in Table~\ref{TabCon}, which aligns each cipher with the attack method that yielded the most favorable trade-off between efficiency and complexity.

\begin{table}
\centering
\begin{tabular}{|c|c|c|}
\hline
\textbf{Cipher} & \textbf{Key size} & \textbf{Most effective (iterations)} \\ \hline
S-DES & 10 & MPS \\ \hline
S-AES & 16 & FQCS \\ \hline
Blowfish & 32 & VQAA \\ \hline
\end{tabular}
\caption{Ciphers analyzed and best hacking methods. The more complex the cipher, the more complex the hacking algorithm.}
\label{TabCon}
\end{table}

Our work opens the door to further research in the study of TN and quantum methods for cybersecurity. For instance, there is room for optimizing the FQCS algorithm to further enhance performance. Some important improvements are utilizing the highly parallelizable nature of the Flexible-PEPS algorithm, which can significantly improve speed, and refining the sampling method to increase efficiency and effectiveness, especially for higher physical bond dimensions. The role of larger bond dimension for large-size keys is also an interesting path to explore. Additionally, this approach can be extended to other protocols like asymmetric-key protocols, such as RSA, and hash functions, such as those used in cryptocurrencies, as shown in Ref. \cite{EnhancedVQAA}. Future work will focus on these aspects to better leverage the potential of TNs in cryptographic key recovery.

{\bf Acknowldgements.-} We acknowledge Donostia International Physics Center (DIPC), Ikerbasque, Basque Government, Diputaci\'on de Gipuzkoa, European Innovation Council (EIC) and Tecnun for constant support, as well as insightful discussions with the teams from Multiverse Computing, DIPC and Tecnun on the algorithms and technical implementations.  This work was supported by the Spanish Ministry of Science and Innovation through the project ``Few-qubit quantum hardware, algorithms and codes, on photonic and solid-state systems''  (PLEC2021-008251) and  by the Diputación Foral de Gipuzkoa through the “Biased quantum error mitigation and applications of quantum computing to gene regulatory networks” project (2024-QUAN-000020). 

{\bf Data availability.-} Te datasets used and/or analysed during the current study are available from the corresponding author on reasonable request.

\bibliographystyle{unsrtnat}
\bibliography{biblio2.bib}

\appendix

\begin{widetext}

\section{Flowchart for FQCS Simulation Algorithm}
\label{app:flowchart}

\noindent In this appendix, we present a detailed flowchart of our Flexible PEPS-based Quantum Circuit Simulator (FQCS). The primary goal of FQCS is to simulate quantum circuits composed of arbitrary one- and two-body gates. The simulation backend is based on the Flexible Projected Entangled Pair States (fPEPS) tensor network algorithm~\cite{patra2024projectedentangledpairstates}.

The fPEPS framework captures entanglement between qubits via virtual bonds connecting vertex tensors that represent the individual qubit states. A higher virtual bond dimension corresponds to higher entanglement content. To maintain computational efficiency—even for non-local gates and densely connected circuits—the method dynamically prunes virtual edges based on the instantaneous correlation structure (quantified by bond entanglement entropy). FQCS leverages this fPEPS formalism to apply gate sequences to a tensor network representing the evolving quantum state. Below, we outline the core steps of the FQCS simulation in a structured flowchart format, encompassing both gate application and entanglement management.

\vspace{1em}
\noindent\textbf{High-Level Flowchart of FQCS (with Hyperparameters $\chi$ and $\kappa$)}
\vspace{0.5em}

\begin{itemize}
    \item \textbf{Initialization:} Initialize a product state on $N$ qubits/spins by assigning a vertex tensor $\Gamma_i$ (a vector) to each site $i$. The initial tensor network contains no virtual bonds. This corresponds to the quantum circuit initialization where all qubits are in the state $|0\rangle_i$.

    \item \textbf{For each gate $O$ in the quantum circuit:}
    \begin{itemize}
        \item \textbf{If $O$ is a one-body gate $O_i$:} \\
        Apply the gate via exact local matrix multiplication:
        \[
            \Gamma_i \leftarrow O_i \cdot \Gamma_i
        \]

        \vspace{0.5cm}
        \item \textbf{If $O$ is a two-body gate $O_{ij}$:}
        \begin{enumerate}
            \item Apply the gate using the \emph{Simple Update} method: apply $O_{ij}$ locally to the vertex tensors $\Gamma_i$ and $\Gamma_j$, followed by singular value decomposition to obtain updated vertex tensors and an edge tensor $\lambda_{ij}$ (or update the existing edge tensor if one already exists).
            \item Truncate the bond dimension of $\lambda_{ij}$ to a maximum value $\chi$, preserving only the dominant singular values.
            \item Update $\Gamma_i$, $\Gamma_j$, and $\lambda_{ij}$ accordingly after truncation.

            \item \textbf{Check:} Does either vertex ($i$ or $j$) now have more than $\kappa$ virtual edges?

            \item \textbf{If yes: For each overconnected vertex $x \in \{i,j\}$}
            \begin{enumerate}
                \item Compute the bond entanglement entropy (BEE) $\mathcal{E}_{xy}$ for all connected edges $(x,y)$.
                \item Sort the BEE values to obtain a ranked list $\bar{\mathcal{E}}_x$.
                \item Identify the edge $(x,y)$ with the lowest BEE to truncate.
                \item \textbf{Truncate edge $(x,y)$ as follows:}
                \begin{itemize}
                    \item Reshape the vertex tensors $\Gamma_x$ and $\Gamma_y$ into matrices $M_x$ and $M_y$.
                    \item Perform a rank-1 truncation (matrix slicing) by retaining only the leading singular value in $\lambda_{xy}$.
                    \item Absorb $\sqrt{\lambda_{xy}}$ into $M_x$ and $M_y$:
                    \[
                        M_x'' = M_x' \cdot \sqrt{\lambda_{xy}}, \quad
                        M_y'' = \sqrt{\lambda_{xy}} \cdot M_y'
                    \]
                    \item Reshape $M_x''$ and $M_y''$ back into updated vertex tensors $\Gamma_x'$ and $\Gamma_y'$.
                    \item Remove the edge $(x,y)$ from the network.
                \end{itemize}
            \end{enumerate}
        \end{enumerate}
    \end{itemize}

    \item \textbf{Repeat:} Iterate over all gates in the circuit.
\end{itemize}

\noindent The final tensor network represents an approximate simulation of the full quantum circuit, governed by the approximation parameters $\chi$ and $\kappa$, where $\chi, \kappa \in \mathbb{Z}^+$. Larger values of these parameters allow more accurate simulations at the cost of increased computational complexity.

\end{widetext}

\color{black}

\end{document}